\documentclass[onecolumn, unsortedaddress, superscriptaddress]{revtex4}
\usepackage{graphicx}
\usepackage{color}
\usepackage{amsmath}

\definecolor{orange}{rgb}{1,0.5,0}

\newcommand{\be}{\begin{equation}}
\newcommand{\ee}{\end{equation}}
\newcommand{\bea}{\begin{eqnarray}}
\newcommand{\eea}{\end{eqnarray}}
\newcommand{\rp}{\right)}
\newcommand{\lp}{\left(}
\newcommand{\rb}{\right]}
\newcommand{\lb}{\left[}

\renewcommand{\vec}[1]{{\bf #1}}

\begin{document}
\title{Photo-excitation Cascade and Multiple Carrier Generation in Graphene}

\author{K.J. Tielrooij}\email[Correspondence to: ]{klaas-jan.tielrooij@icfo.es,   frank.koppens@icfo.es}
\affiliation{ICFO - Institut de Ci\'encies Fot\'oniques,
Mediterranean Technology Park, Castelldefels (Barcelona) 08860,
Spain}
\author{J.C.W. Song}
\affiliation{Department of Physics, Massachusetts Institute of
Technology, Cambridge, Massachusetts 02139, USA}
\affiliation{School of Engineering and Applied Sciences, Harvard
University, Cambridge, Massachusetts 02138, USA}
\author{S.A. Jensen}
\affiliation{Max Planck Institute for Polymer Research,
Ackermannweg 10, 55128 Mainz, Germany} \affiliation{FOM Institute
AMOLF, Amsterdam, Science Park 104, 1098 XG Amsterdam,
Netherlands}
\author{A. Centeno}
\affiliation{Graphenea SA, 20018 Donostia-San Sebasti\'an, Spain}
\author{A. Pesquera}
\affiliation{Graphenea SA, 20018 Donostia-San Sebasti\'an, Spain}
\author{A. Zurutuza Elorza}
\affiliation{Graphenea SA, 20018 Donostia-San Sebasti\'an, Spain}
\author{M. Bonn}
\affiliation{Max Planck Institute for Polymer Research,
Ackermannweg 10, 55128 Mainz, Germany}
\author{L.S. Levitov}
\affiliation{Department of Physics, Massachusetts Institute of
Technology, Cambridge, Massachusetts 02139, USA}
\author{F.H.L. Koppens}\email[Correspondence to: ]{klaas-jan.tielrooij@icfo.es,   frank.koppens@icfo.es}
\affiliation{ICFO - Institut de Ci\'encies Fot\'oniques,
Mediterranean Technology Park, Castelldefels (Barcelona) 08860,
Spain}

\newpage

\begin{abstract}
The conversion of light into free electron-hole pairs constitutes
the key process in the fields of photodetection and photovoltaics.
The efficiency of this process depends on the competition of
different relaxation pathways and can be greatly enhanced when
photoexcited carriers do not lose energy as heat, but instead
transfer their excess energy into the production of additional
electron-hole pairs via carrier-carrier scattering processes. Here
we use Optical pump - Terahertz probe measurements to show that in
graphene carrier-carrier scattering is unprecedentedly efficient
and dominates the ultrafast energy relaxation of photoexcited
carriers, prevailing over optical phonon emission in a wide range
of photon wavelengths. Our results indicate that this leads to the
production of secondary hot electrons, originating from the
conduction band. Since hot electrons in graphene can drive
currents, multiple hot carrier generation makes graphene a
promising material for highly efficient broadband extraction of
light energy into electronic degrees of freedom, enabling
high-efficiency optoelectronic applications.
\end{abstract}

\maketitle

\section{Main text}

For many optoelectronic applications, it is highly desirable to
identify materials in which an absorbed photon is efficiently
converted to electronic excitations. The unique properties of
graphene, such as its gapless band structure, flat absorption
spectrum \cite{Nair2008} and strong electron-electron interactions
\cite{Kotov2012}, make it a highly promising material for
efficient broadband photon-electron conversion
\cite{Bonaccorso2010}. Indeed, recent theoretical work has
anticipated that in graphene multiple electron-hole pairs can be
created from a single absorbed photon during energy relaxation of
the primary photoexcited e-h pair \cite{Winzer2010,Winzer2012}. A
photo-excited carrier relaxes initially trough two competing
pathways: carrier-carrier scattering and optical phonon emission.
In the former process the energy of photoexcited carriers remains
in the electron system, being transferred to secondary electrons
that gain energy (become hot), whereas in the phonon emission
process the energy is lost to the lattice as heat. While recent
experiments have shown that photoexcitation of graphene can
generate hot carriers \cite{Gabor2011,Song11}, it remains unknown
how efficient this process is with respect to optical phonon
emission.
\\

Here we study the energy relaxation process of the primary
photoexcited e-h pair in doped single-layer graphene. In
particular, we quantify the branching ratio between the two
competing relaxation pathways. Given the challenging timescale
with which these processes occur, we employ an ultrafast Optical
pump - Terahertz (THz) probe measurement technique, where we
exploit the variation of the photon energy of the pump light.
Changing this photon energy is crucial as it allows us to prepare
the system with photoexcited carriers having a prescribed initial
energy determined by the photon energy, and follow the ensuing
energy relaxation dynamics. We show experimentally, in combination
with theoretical modeling, that carrier-carrier scattering is the
dominant relaxation process. This process leads to the creation of
secondary hot electrons that originate from the conduction band.
\\

We note that assigning a `conventional name' to a process in which
secondary hot carriers are generated by photoexcited carriers in
graphene, is by no means a trivial matter. This is so because
somewhat different nomenclature is used in the optical studies of
semiconductors and metals: electrons and holes in semiconductors
are defined with respect to the conduction and valence bands,
whereas in metals the distinction between the states above and
below the Fermi level plays the key role. Doped graphene can be
viewed as a mixture of both: it is a semimetal with the Fermi
level detuned away from the Dirac point. To minimize confusion,
and at the same time to make the discussion of our results
unambiguous, we will denote the process in question as
``hot-carrier multiplication.'' This \textit{intraband} process is
different from conventional carrier multiplication observed in
semiconductor systems
\cite{Pijpers2007,Pijpers2009,Schaller2004,Schaller2005} and
theoretically predicted for undoped graphene \cite{Winzer2010,
Winzer2012}, where additional e-h pairs originate from \textit{
interband} transitions. However, the generation of secondary hot
carriers from the conduction band in doped graphene is a
technologically relevant relaxation process since the
thermoelectric effect dominates the optoelectronic response of
graphene \cite{Gabor2011,Song11}. For hot carrier multiplication,
the total number of carriers in the conduction band does not
change due to carrier-carrier scattering. However, the number of
hot carriers (i.e.\ carriers with an energy above the Fermi level)
increases. It is via this multiplication of hot electrons in the
conduction band that the energy of the primary photoexcited
carrier is ``harvested'' by the electron subsystem and later used
to generate an optoelectronic response.
\\

The employed technique consists of an ultrafast optical pump pulse
that excites carriers and a THz probe pulse that passes through
the sample after a variable delay time. The THz pulses afford an
exquisite time-resolved probe of the high frequency
photoconductive response of photoexcited carriers, as reviewed in
Ref.\ \cite{Ulbricht2011}. This technique has been used before
with a fixed pump wavelength to study charge dynamics in
multilayer graphene \cite{george, Breusing2009, Breusing2011,
kampfrath05,Winnerl2011, Strait2011}, and with a variable pump
wavelength to study the effects of carrier-carrier interaction in
semiconductor materials \cite{Pijpers2009, Pijpers2007,
Schaller2005}. Here we apply this technique with variable pump
wavelength to examine the energy relaxation cascade of
photoexcited carriers in graphene. We use a monolayer of
intrinsically doped graphene, in contrast to previous optical pump
- THz probe studies which used multilayer (undoped) graphene
\cite{george,Breusing2009, Breusing2011, kampfrath05,Winnerl2011,
Strait2011}. We find that the photoexcited density of carriers
with energy above the Fermi energy scales linearly with photon
energy (for constant absorbed photon density). This scaling is
found over a wide range that spans almost an order of magnitude in
photon wavelength. Our experimentally observed linear scaling
indicates that carrier-carrier scattering is remarkably efficient,
in excellent agreement with our results from a theoretical model
that considers electron-electron scattering and electron-optical
phonon scattering.
\\

The intrinsically doped graphene sample that we use for our study
consists of a monolayer of CVD graphene transferred onto a quartz
substrate. From Raman spectroscopy we estimate a Fermi energy of
$\mu \sim 0.17 \pm 0.05$ eV, which corresponds to an intrinsic
carrier concentration of $\sim$ 2$\times$10$^{12}$
carriers/cm$^2$. We further characterize the sample using THz
transmission (without optical excitation) and find that the
graphene monolayer has a spectrally flat absorption of $\sim\,
5\%$ in the 0.4--1.6 THz region. This absorption is due to
intraband momentum scattering of the intrinsic carriers (Drude
conductivity) with an extracted average transport time of
$\tau_{\rm tr} \sim 20$ fs (see supplementary online material for
a detailed sample characterization).
\\

Our method for probing the energy relaxation dynamics of
photoexcited carriers is illustrated in Fig.\ \textbf{1a}. We
optically excite the graphene/quartz stack and examine the
pump-induced change in THz transmission $\Delta T = T - T_0$, were
$T$ and $T_0$ are the transmission with and without
photoexcitation, respectively. The THz pulses follow the pump
pulses after a tunable time delay with a time resolution of
$\sim$100 fs. Because the graphene layer is thin, the change in
transmission is directly related to the photoconductive response
so that $\frac{\Delta T}{T_0} \propto -\Delta \sigma$ (see
supplementary online material). Here $\Delta \sigma$ is the
pump-induced change in the THz conductivity, which we refer to as
the THz photoconductivity.
\\

The essential features of the observed dynamics are exemplified by
two typical differential transmission time traces shown in Fig.\
\textbf{1b}. The observed temporal evolution is non-monotonic: the
differential transmission first rises in an approximately linear
fashion, reaching a peak at a delay time $\tau_{\rm peak} \sim
200$ fs, and finally decays on a longer timescale $\sim 1.4$ ps.
These observations can be explained as follows. During the initial
rise, $t\lesssim \tau_{\rm peak}$, carrier-carrier scattering
between the photoexcited carriers and the carriers in the Fermi
sea promotes carriers from below to above the Fermi level (see
middle inset of Fig.\ \textbf{1b}). Thus the effective temperature
of the carrier distribution increases, peaking at
$t\approx\tau_{\rm peak}$ when the photoexcited carriers have
relaxed and a hot carrier distribution is established. Subsequent
relaxation occurs due to electron-lattice cooling, which is a
relatively slow process with a characteristic timescale ($\sim
1.4$ ps) much greater than $\tau_{\rm peak}$. These dynamics are
very similar to those observed in earlier optical pump-THz probe
studies on multilayer (undoped) graphene \cite{george,
Breusing2009, Breusing2011, kampfrath05,Winnerl2011, Strait2011}.
\\

The measured change in transmission $\frac{\Delta T}{T_0}$ is
positive, meaning that the conductivity is reduced as a result of
photoexcitation, $\Delta\sigma<0$. The observation of negative
photoconductivity is in agreement with recent THz pump-probe
studies on monolayer graphene \cite{Hwang2011, frenzel2012}. The
reduction in conductivity is naturally related to secondary hot
carrier excitation processes: Since the momentum scattering time
$\tau_{\rm tr} (\epsilon)$ increases with carrier energy
$\epsilon$ \cite{nomura,ando,dassarmareview}, the creation of a
hot carrier distribution after photoexcitation leads to a change
in the real part of conductivity that has a negative sign (see
supplementary online text). In contrast, for multilayer (undoped)
graphene a positive change in the conductivity was observed,
because in that case photoexcitation leads to additional e-h pairs
in the conduction band \cite{george, Breusing2009, Breusing2011,
kampfrath05,Winnerl2011, Strait2011}.
\\

Key insight into the processes contributing to the energy
relaxation cascade comes from examining how the differential
transmission signal $\frac{\Delta T}{T_0}$ peak value scales with
the photon energy $hf$, which in our experiment is varied over a
wide range, from the infrared ($0.16$ eV) to the ultraviolet
($4.65$ eV). Since the initial photoexcited carrier energy,
$\epsilon_i = \frac{1}{2} hf$, determines where the cascade
begins, we can track how the photoexcited carriers relax at each
stage of the energy re-distribution process. As a first step, we
analyze the peak value $\frac{\Delta T}{T_0}$ dependence on
absorbed photon density $N_{\rm photon}$ (i.e.\ the number of
absorbed photons per unit of area), which is displayed in Fig.\
\textbf{1c}, for six different photon wavelengths. In the fluence
regime employed here, the signal increases linearly with absorbed
photon density for each photon energy. The linear dependence of
THz photoconductivity on fluence at fixed photon energy indicates
that in this regime each photoexcited carrier acts independently
from the other photoexcited carriers.
\\

Proceeding with the analysis we observe that increasing the photon
energy at a {\it fixed absorbed photon density} leads to a larger
differential transmission signal at the peak. This is clear from
the slopes in Fig.\ \textbf{1c} that increase with photon energy.
The origin of the increased signal for increased photon energy (at
fixed absorbed photon density) is shown schematically in Fig.\
\textbf{1d}. Here, increasing the photon energy leads to an
increased number of electron-electron scattering events during the
relaxation cascade and thus a hotter carrier distribution. In
Fig.\ \textbf{2a} we show the effect of increasing the photon
energy by plotting the peak differential transmission signal
normalized by absorbed photon density. Notably, the normalized
signal scales approximately linearly with the photon energy,
whereas energy relaxation through phonon emission would lead to a
normalized signal that would be independent of photon energy.
\\

It is instructive to combine these observations in a unified
picture which provides an intuitive ``bird's eye view'' of the
energy relaxation cascade. We do this by plotting the interpolated
experimental contours for the  $\Delta T/T_0$ peak value as a
function of photon energy and photon number (see Fig.\
\textbf{2d}). Strikingly, the contours of constant $\Delta T/T_0$
bunch up at high photon energy and spread out at low photon
energy. This confirms that the two ways to achieve a hotter
distribution of carriers -- either by increasing the absorbed
photon density or by increasing the photon energy -- are
completely interchangeable. The hyperbolical shape also indicates
that the differential transmission signal scales with the fluence
(incident energy per area, $N_{\rm photon} \times hf$). This
constitutes a clear qualitative signature of the dominance of
carrier-carrier scattering. Without carrier-carrier scattering,
the magnitude of the response would be determined only by the
absorbed photon density, and not the excitation energy; the
contour lines would have been essentially vertical, with no change
in $\Delta T/T_0$ as photon energy is varied (see also Fig.\
\textbf{2f}). From these ``bird's eye view'' plots we conclude
that carrier-carrier scattering plays an important role in the
energy relaxation cascade. Below we develop this notion more
quantitatively and estimate  the (energy dependent) efficiency of
carrier-carrier scattering in graphene.
\\

The efficiency of carrier-carrier scattering depends on the
branching ratio between the two processes: (i) carrier-carrier
scattering vs. (ii) electron-optical phonon scattering  that occur
during the rise stage, $0<t\lesssim \tau_{\rm peak}$. To extract
this branching ratio, we develop a simple model for energy
relaxation in the photoexcitation cascade and compare it with the
data. The relaxation can be described in a general form via
$d{\epsilon}/dt=-\mathcal{J}_{\rm el-el}(\epsilon)  -
\mathcal{J}_{\rm el-ph}(\epsilon)$, where $\epsilon$ is the
photoexcited carrier energy and $\mathcal{J}_{\rm el-el} $ and
$\mathcal{J}_{\rm el-ph}$ represent the energy relaxation rates
for processes (i) and (ii). Our analysis relies on the fact that
the characteristic time of the photoexcitation cascade starting at
the photon energy $\epsilon = hf/2$ (at $t=0$) and ending at the
Fermi energy $\epsilon \approx \mu$ (at $t=\tau_{\rm peak}$) is
much longer than the carrier-carrier scattering time. Indeed,
typical values $\tau_{\rm peak} \sim 200$ fs measured in our
experiment (see Fig.\ \textbf{1b}) are considerably longer than
the reported values for carrier-carrier scattering times, which
are well below 100 fs \cite{george,song12}. Thus the electron
subsystem during the cascade can be described using an effective
electron temperature which is distinct from the lattice
temperature. Thermal equilibration with the lattice in graphene is
slow \cite{macdonald,wong09,song12a}, and in our sample takes
$\sim$ 1.4 ps (independent of photon energy and fluence in the
regime considered here). This separation of time scales allows us
to describe the electronic system using the electron temperature
approximation.
\\

Carrier-carrier scattering in graphene leads to the creation of
secondary hot electrons that originate from the conduction band.
These secondary hot electrons give a negative contribution to the
THz photoconductivity, since the momentum scattering time
$\tau_{\rm tr} (\epsilon)$ increases with carrier energy
$\epsilon$ \cite{nomura,ando,dassarmareview}. Accounting for fast
thermalization of the secondary carriers, the net change in the
THz photoconductivity can be expressed as $-2a N_{\rm
photon}\mathcal{J}_{\rm el-el}$, where $N_{\rm photon}$ is the
absorbed photon density, and the pre-factor $a$ is estimated in
the supplementary online material (the factor of 2 accounts for
two carriers (electron and hole) produced per photon). Integrating
over the cascade, we obtain the conductivity at the peak:

\be \label{eq:photocontot} \Delta \sigma \approx -2 N_{\rm photon}
\int_{t_0}^{t_0+\tau_{\rm peak}} dt  a\mathcal{J}_{\rm el-el} = -2
N_{\rm photon} \int^{hf/2}_0 d\epsilon \frac{ a\mathcal{J}_{\rm
el-el}(\epsilon)}{\mathcal{J}_{\rm el-el}(\epsilon)  +
\mathcal{J}_{\rm el-ph} (\epsilon)} . \ee
\\

In the case that electron-electron scattering processes dominate,
$\mathcal{J}_{\rm el-el} \gg \mathcal{J}_{\rm el-ph}$, Equation
(\ref{eq:photocontot}) directly leads to linear scaling. In this
case, since $a$ is approximately energy independent (see
supplementary online material), we have $\Delta \sigma \approx -a
N_{\rm photon}hf$, i.e.\ perfectly linear scaling. In a more
realistic regime, when $\mathcal{J}_{\rm el-el} \sim
\mathcal{J}_{\rm el-ph}$, the THz photoconductivity in Equation
(\ref{eq:photocontot}) is sensitive to the branching ratio of the
two processes, $\alpha(\epsilon) = \mathcal{J}_{\rm el-el} /
\mathcal{J}_{\rm el-ph}$. Using $\mathcal{J}_{\rm el-el}$ from
electron-electron scattering events described in Ref.
\cite{song12} and writing $\mathcal{J}_{\rm el-ph} = \gamma
(\epsilon-\omega_0) \Theta ( \epsilon - \omega_0 - E_F)$ we obtain
the branching ratios plotted in Fig.\ \textbf{2b}. Here $\omega_0$
is the optical phonon energy (0.2 eV), $\Theta$ is a step function
and $\gamma$ (in ${\rm (ps)}^{-1}$) is the coupling constant
between electrons and optical phonons, which we use as a fitting
parameter to obtain the branching ratio. Since other Auger
processes like interband relaxation of carriers are blocked
kinematically \cite{Foster2009}, $J_{\rm el-el}$ from impact
excitation captures the relevant carrier-carrier scattering
processes. Since the branching ratio has a strong energy
dependence (approximately $\propto 1/\epsilon$ at high energies),
we use the scaling of $\Delta \sigma$ with photon energy as a
sensitive probe of the magnitude of the branching ratio. Indeed,
$\Delta \sigma /N_{\rm ph}$ from Equation (1) begins to deviate
from linear scaling (with $hf$) significantly when the branching
ratio becomes smaller than unity, $\alpha (\epsilon) < 1$.
\\

In Fig.\ \textbf{2a}, we compare Equation (1) for three values of
the electron-phonon coupling constant $\gamma$ to our observed
differential transmission signal normalized by absorbed photon
density (see supplementary online material). The best-fit value
$\gamma \approx 4 \, {\rm ps}^{-1}$ corresponds to the solid curve
in Fig.\ \textbf{2a}. The branching ratio values indicate that
electron-electron scattering dominates the energy relaxation
cascade. For stronger electron-phonon coupling, i.e.\ larger
values of $\gamma$, we find that $\Delta \sigma$ from Equation (1)
bends down at higher energies, significantly deviating from
linearity, as illustrated by the dotted curve for $\gamma = 10\,
{\rm ps}^{-1}$. We can exclude fits with $\gamma > 10 {\rm
ps}^{-1}$ as the data clearly lie in the range $ 1\, {\rm
ps}^{-1}\lesssim\gamma\lesssim 10 \, {\rm ps}^{-1}$. Inspecting
the branching ratios that correspond to $\gamma = 1$, $4$, and $10
{\rm ps}^{-1}$ (see Fig.\ \textbf{2b}) we conclude that the energy
relaxation cascade that produces our observed $\Delta T/T_0$ have
branching ratio $\alpha(\epsilon) \gtrsim 1$. Hence,
electron-electron scattering dominates the energy relaxation
cascade of photoexcited carriers, prevailing over electron-optical
phonon scattering.
\\

We find good agreement between our experimental observations and
values predicted by the model. First of all, the branching ratio
we find in Fig.\ \textbf{2b} is in good agreement with the
theoretically predicted branching ratio. Using the known value of
the electron-optical phonon deformation potential
\cite{macdonald}, we compute $\gamma = 1.36\,  {\rm ps}^{-1}$ (see
supplementary online material), which lies in the range of
$\gamma$'s obtained by fitting the data. In the supplementary text
we also show that the experimentally observed magnitude of the
signal is in reasonable agreement with the theoretical result for
$a$ based on the effective temperature model. Furthermore, it is
interesting to note that the rise time of the pump probe signal in
Fig.\ \textbf{1b} -- corresponding to the time needed for the
energy relaxation cascade -- is longer for excitation with a
higher photon energy. Although our experimental time resolution is
only marginally smaller than the rise time, the variation with
photon energy indicates that the observed behavior at $t\lesssim
\tau_{\rm peak}$ is not limited by time resolution. The observed
dependence is in qualitative agreement with the theoretical
prediction of longer cascade times at higher photon energy, where
more carrier-carrier scattering events occur during the cascade
\cite{song12}.
\\

Using the branching ratios plotted in Fig.\ \textbf{2b}, we
calculate the fraction of the photon energy that remains in the
electronic system after the cascade (integrated e-e efficiency) as
\be \eta(\epsilon_i) = \frac{1}{\epsilon_i}\int_0^{\epsilon_i}
\frac{\alpha(\epsilon)}{\alpha(\epsilon)+1}d\epsilon
\hspace{1.5cm} \epsilon_i = hf/2  .
 \ee

We plot $\eta(\epsilon_i)$ for $\gamma = 1 \, {\rm ps}^{-1}-10 \,
{\rm ps}^{-1}$ in Fig.\ \textbf{2c} showing that more than 50\% of
the photon energy remains in the electronic system even for
photoexcitation energies as high as $hf = 3 \, {\rm eV}$. This
shows that carrier-carrier interaction in graphene is highly
efficient. We note that our model may overestimate the efficiency
at low energies, as close to the Fermi surface energy relaxation
may depend on pathways that were not included in the model:
acoustic phonons, flexural phonons and substrate surface phonons.
However, since these processes only become important close to the
Fermi surface we expect their impact on the total efficiency to be
small.
\\

For most applications the relevant figure of merit is the
integrated efficiency $\eta$ as it describes the fraction of light
energy that is passed to the electronic system, where hot carriers
can drive currents of optoelectronic systems. However it is also
interesting to examine how many hot electrons are created from a
single incident photon. Our model predicts that the number of
secondary hot electrons scales approximately linearly with photon
energy, $N\sim\epsilon_i/\mu$ \cite{song12}. Taking into account
the extracted efficiency of 80\% for excitation with 3 eV pump
light, we find that 9 additional hot electrons are created, that
get promoted from below to above the Fermi level in the conduction
band. \cite{song12}.
\\

Our study reveals efficient carrier-carrier interaction in
graphene, thereby resolving a long-standing question about the
relative importance of electron-electron scattering versus
emission of optical phonons in the energy relaxation cascade
triggered by photoexcitation. Crucially, the transfer of energy
from photoexcited carriers to electronic degrees of freedom in
graphene is efficient over a wide range of frequencies (from the
UV to the infrared), unlike conventional semiconductor systems
where the frequency range is limited by the band gap. Furthermore,
the number of secondary hot electrons is expected to be highly
sensitive to the doping level \cite{song12}, enabling effective
manipulation of the energy cascade pathways. Thus graphene enables
enhanced quantum efficiencies and tunable energy transfer over a
wide spectral range.

\newpage

\bibliographystyle{main-bib}

\section{Acknowledgements}

We acknowledge financial support from an NWO Rubicon grant (KJT),
the NSS program, Singapore (JS), the Office of Naval Research
Grant No. N00014-09-1-0724 (LL), and Fundacio Cellex Barcelona and
ERC Career integration grant GRANOP (FK).

\newpage

\section{Figures}

\begin{figure} [h!!!!!]
   \centering
   \includegraphics [scale=0.75]
   {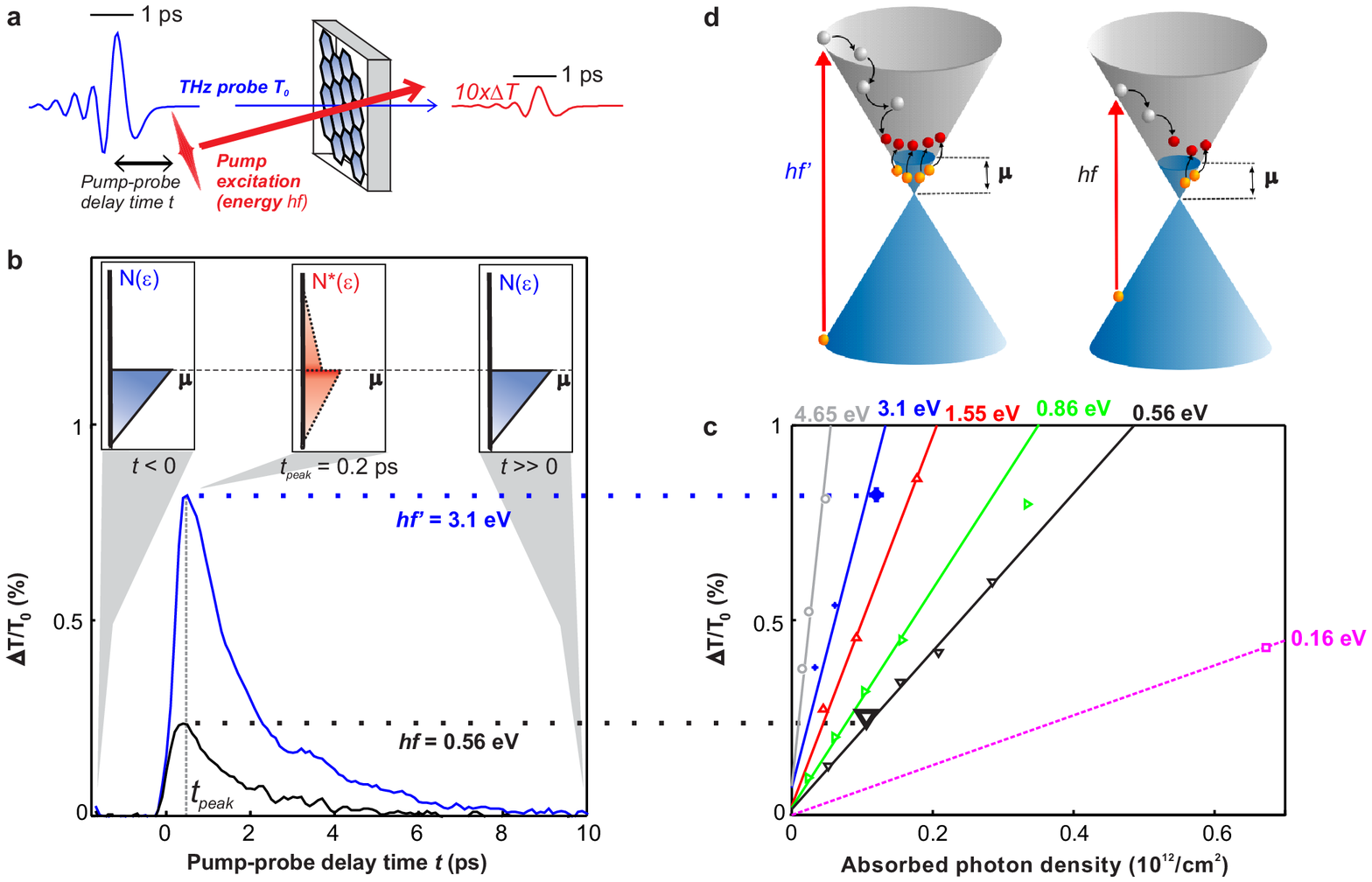} 
   \caption{}
   \label{F_pump_probe}
\end{figure}

\begin{figure} [h!!!!!]
   \centering
   \includegraphics [scale=0.85]
   {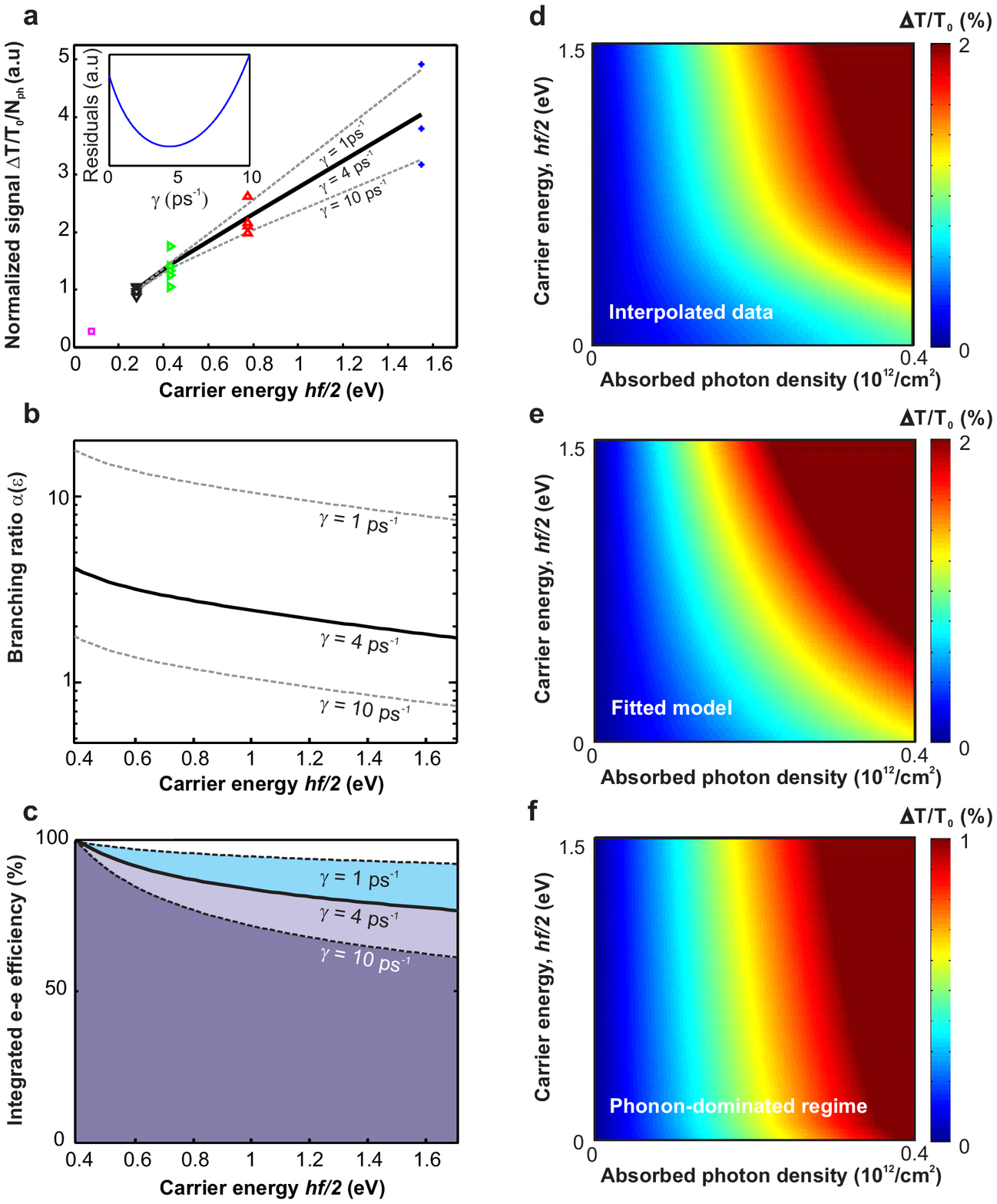}
   \caption{}
   \label{F_peaks}
\end{figure}

\clearpage

\section{Figure captions}

\large \textbf{Figure 1} \normalsize

\textit{Experimental realization and results.} \textbf{a)}
Experimental observation of carrier dynamics. Photoexcitation of
electron-hole pairs in graphene monolayer by a pump pulse is
followed by a Terahertz probe pulse that provides a measurement of
the pump-induced change in THz electric field transmission $\Delta
T = T - T_0$ (red curves). The time dynamics of the photoexcited
carriers is studied by varying the time delay $t$ between the pump
pulse and the probe pulse. The change in transmission $\Delta T $,
which is directly proportional to the change in THz conductivity,
is used to characterize the degree to which secondary hot
electrons are produced as the photoexcited electron cascades down
to the Fermi level. \textbf{b)} Time resolved carrier dynamics for
two different photon energies: the pump-induced change in THz
transmission $\Delta T/T_0$ as a function of pump-probe delay time
$t$ for fixed absorbed photon density $N_{\rm photon} \sim 0.1
\times 10^{12}$ cm$^{-2}$. The insets show a schematic
representation of the carrier distribution $N(\epsilon)$ before
($t < 0$) and long after ($t \gg \tau_{\rm peak}$) pump
excitation, and the hottest carrier distribution $N^*(\epsilon)$
directly after the energy relaxation cascade ($t = \tau_{\rm
peak}$). \textbf{c)} Scaling of the differential transmission
signal $\frac{\Delta T}{T_0}$ peak values, obtained from the time
traces as in \textbf{b} at $t = \tau_{\rm peak}$ for six photon
energies as a function of absorbed photon density $N_{\rm
photon}$. The lines are linear fits to guide the eye, showing that
the signal increases linearly with absorbed photon density. For a
given absorbed photon density $N_{\rm photon}$ a higher photon
energy $hf$ leads to an increased signal, corresponding to a
hotter carrier distribution. \textbf{d)} The effect of varying
$hf$ on the energy cascade illustrated for two photon energies,
$hf'>hf$. Photoexcitation creates a primary electron-hole pair,
and triggers a cascade of carrier-carrier scattering steps, where
energy is transferred to multiple secondary hot electrons in the
conduction band, generating a hot carrier distribution. The number
of secondary hot electrons increases with photon energy, leading
to a hotter carrier distribution and a larger observed
$\frac{\Delta T}{T_0}$ signal.

\large \textbf{Figure 2} \normalsize

\textit{Carrier-carrier scattering efficiency.} \textbf{a)}
Extraction of the branching ratio between e-e scattering and
optical phonon emission from comparison of the experimental data
and model: The pump-probe signal $\frac{\Delta T}{T_0}$ peak value
normalized by absorbed photon density $N_{\rm photon}$ features
approximately linear scaling with $hf$, indicating that
electron-electron scattering dominates the energy relaxation
cascade (see text). Deviation from linearity is accounted for in
the model by including optical phonon emission with a coupling
strength $\gamma$ (see Equation (1)). The best-fit curve ($\gamma
= 4\,{\rm ps}^{-1}$, solid line) and the curves that marginally
agree with the experimental data ($\gamma = 1\,{\rm ps}^{-1}$ and
$\gamma = 10\,{\rm ps}^{-1}$, dashed line) are shown. The inset
shows the fit residuals. \textbf{b)} The branching ratio for the
carrier-carrier scattering and optical phonon emissions pathways
as a function of initial carrier energy for the three coupling
strengths shown in \textbf{a}. Larger-than-one values indicate
that electron-electron scattering is the dominant
energy-relaxation pathway. \textbf{c)} The integrated efficiency
for the carrier-carrier scattering pathway as a function of
carrier energy for the same three coupling strength values:
best-fit (solid line), and upper and lower bounds. The estimated
efficiencies are well above 50\%, indicating that a large fraction
of incident photon energy is transferred to the electronic system
through efficient carrier-carrier scattering. \textbf{d,e,f)}
``Bird's eye view'' of the differential signal vs. photon energy
and absorbed photon number density for experimental data ({\bf
d}), model best-fit ({\bf e}) and (simulated)
phonon-emission-dominated cascade  ({\bf f}). For efficient
carrier-carrier scattering, the differential transmission signal
is expected to increase with both absorbed photon density $N_{\rm
photon}$ and with carrier energy $hf/2$. The contour lines of
constant $\frac{\Delta T}{T_0}$ as a function of $N_{\rm photon}$
and $hf/2$ obtained by interpolation of the data (\textbf{d})
indeed show this behavior. The good overall correspondence with
the model for a best-fit $\gamma$ value (\textbf{e}) further
corroborates the conclusion that e-e scattering dominates the
cascade. In contrast, simulation of phonon-emission dominated
cascade ($\gamma = 100$ ps$^{-1}$, \textbf{f}) shows significant
departure from the data.

\clearpage

\section{Appendix}

\renewcommand{\thesection}{S\arabic{section}}
\renewcommand{\thefigure}{S\arabic{figure}}
\renewcommand{\theequation}{S\arabic{equation}}
\setcounter{figure}{0} \setcounter{section}{0}
\setcounter{equation}{0}

\renewcommand{\vec}[1]{{\bf #1}}
\renewcommand{\Im}{\rm Im\,}
\renewcommand{\Re}{\rm Re\,}

\section{Optical pump - Terahertz probe technique}

The optical pump - Terahertz probe setup is based on an ultrafast
amplified laser system with a pulse duration of $\sim$50
femtoseconds, a center wavelength of 800 nm and a repetition rate
of 1 kHz. Part of the output is used to create the optical pump
pulses and part is used to create the THz probe pulses. The
optical pump pulses have a photon energy that we vary using
non-linear optical conversion. To create 400 nm excitation pulses,
we use frequency doubling in a $\beta$-Barium Borate (BBO)
nonlinear crystal ($\theta$ = 29.2$^\circ$, 1 mm) and for the 267
nm pulses, we use the sum frequency of 400 nm and 800 nm pulses in
a second BBO crystal ($\theta$ = 44.3$^\circ$, 0.2 mm). For the
excitation pulses in the infrared, we feed the 800 nm pulses into
a home-built optical parametric amplifier (OPA), which we tune to
either produce signal pulses at 1400 nm or idler pulses at 2100
nm. Finally, for the 8 $\mu$m pulses, we use a AgGaS$_2$ nonlinear
crystal ($\theta$ = 50$^\circ$, 1.2 mm) for difference frequency
mixing between 1455 nm idler pulses and 1778 nm idler pulses. The
pump pulses pass through a diffuser to ensure spatially
homogeneous excitation of the graphene sample and through a 500 Hz
chopper so that we alternately probe the THz electric field
transmission with and without photo-excitation by pump pulses,
$T_0 + \Delta T$ and $T_0$, respectively. We determine the pump
fluence at each wavelength by measuring the transmission trough
five holes of different well-known sizes with a calibrated power
meter. The 8 $\mu$m and 267 nm pump pulses had very low fluences
and therefore did not pass through the diffuser. Excitation
without diffuser could lead to an underestimation of the absorbed
photon density in the case of hot spots in the beam profile and
therefore these two photon energies are not taken into account for
the determination of the efficiency of carrier multiplication.
\\

The THz probe pulses are created through optical rectification in
ZnTe nonlinear crystal (110 orientation, 0.5 mm) and focused on
the sample. After recollecting and collimating, the THz beam is
focused on a second ZnTe crystal and overlapped in time with an
800 nm sampling beam, whose polarization is rotated by the
presence of the electric field of the THz beam. We then detect
this polarization rotation, which is directly proportional to the
electric field strength of the THz beam. By varying the time delay
between the THz beam and the 800 nm sampling beam, we obtain the
time evolution of the THz pulse. For the pump-probe measurements
described here, we position this time delay such that we are at
the peak of the THz pulse. We then change the time delay of the
pump pulses with respect to the peak of the THz pulse in order to
follow the temporal evolution of the THz pump-probe signal, which
is proportional to the THz photoconductivity in the thin film
approximation: $\Delta \sigma = -{{\Delta T}\over{T_0}} {{n_{\rm
s} + 1}\over{Z_0}}$. Here $n_{\rm s} = 1.95$
\cite{Grischkowsky1990} is the refractive index of the quartz
substrate and $Z_0 = 377$ $\Omega$ is the free space impedance.
All measurements were done at room temperature. We verify the sign
of the pump-probe signal by comparing measuring the signal through
a silicon sample and our monolayer graphene sample sequentially
without changing the chopper in Lock-in Amplifier phase settings.
\\

\section{Sample characterization}
\label{S:sample_characterization}

\begin{figure} [h!!!!!]
   \centering
   \includegraphics [scale=0.7]
   {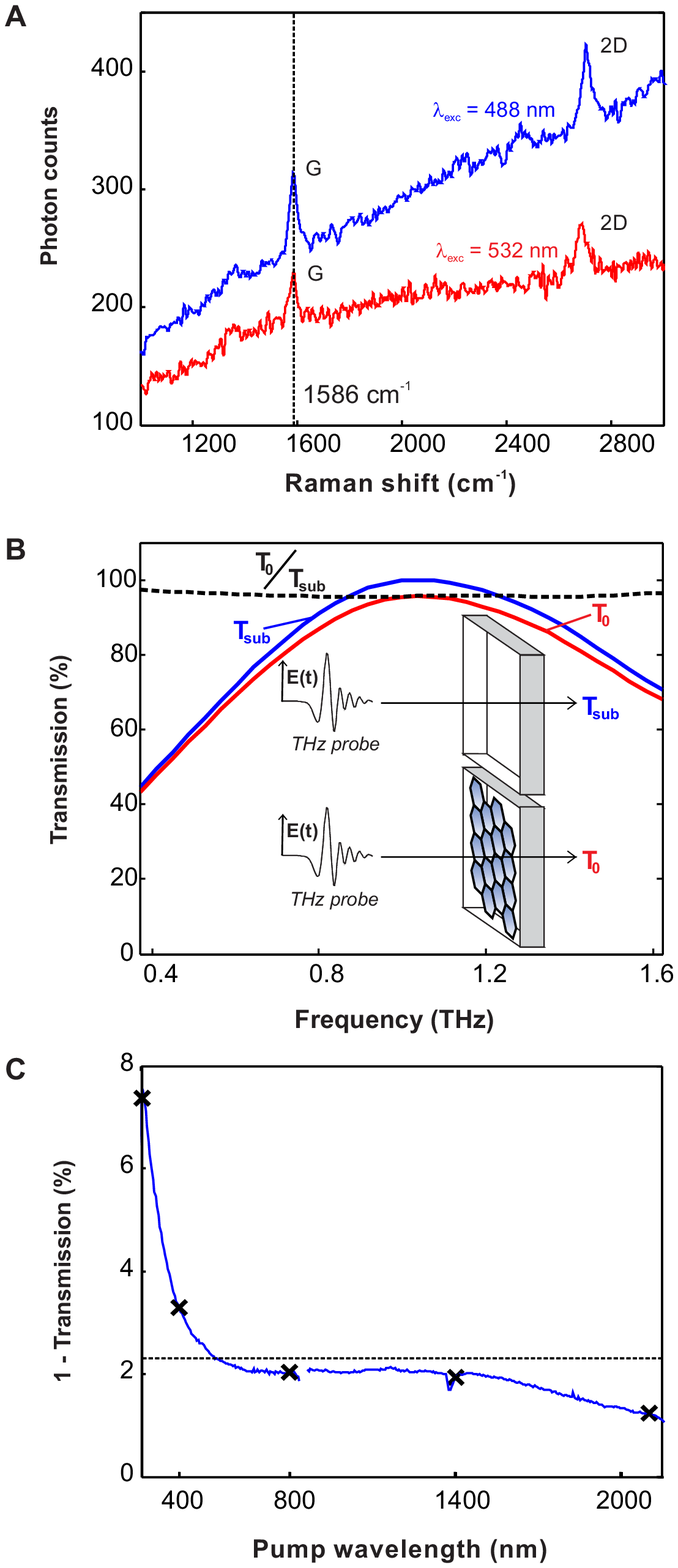}
   \caption{\textit{Sample characterization.} \textbf{A)} Raman spectroscopy to determine the
   intrinsic carrier density through substrate doping. \textbf{B)} Terahertz transmission spectrum, showing
   spectrally flat absorption of $\sim$5\%, due to the conductivity of the intrinsic free carriers.
   \textbf{C)} Absorption spectrum to determine the absorption for
   the used pump wavelengths, indicated by black crosses.}
   \label{F_SOM_abs_raman}
\end{figure}

The graphene samples consist of a large ($\sim1$ cm$^2$) single
graphene layer that was grown by chemical vapor deposition (CVD)
and subsequently transferred onto a 200 $\mu$m thick fused silica
substrate (CFS-2520 from UQG Optics). To determine the intrinsic
doping of the graphene layer, we performed Raman spectroscopy (see
Fig.\ \ref{F_SOM_abs_raman}A). We find a G peak located at 1586
cm$^{-1}$ corresponding to a Fermi level of $E_F = 0.17 \pm 0.05$
eV and thus an intrinsic doping of $n_{\rm intr} \sim 2 \times
10^{12}$ carriers/cm$^{-2}$ \cite{Pisana2007,Das2008}. The FWHM of
the 2D peak is 30$\pm5$ cm$^{-1}$, confirming that our sample is
monolayer graphene. We measured the THz transmission (without
optical pump excitation) and find that the graphene layer leads to
a spectrally flat absorption of $\sim$ 5\% in the THz region (see
Fig.\ \ref{F_SOM_abs_raman}B). Using the thin film approximation
we find that this absorption corresponds to an intrinsic sheet
conductivity of $\sigma_{\rm intr} \approx {{T_{\rm sub} -
T_0}\over{T_{\rm sub}}} {{n_{\rm s} + 1}\over{Z_0}} = $ 4$\times
10^{-4}$ S/m (or $\sim 10 e^2/h$), with $T_{\rm sub}$ the
transmission through the substrate without graphene. The (Drude)
sheet conductivity in graphene is determined by the carrier
density and the weighted momentum scattering time $\langle
\tau_{\rm tr} \rangle$. We use the carrier density from the Raman
measurements to extract a momentum scattering time of $\sim$20 fs,
similar to earlier conductivity studies of large area CVD graphene
\cite{Horng2011,Strait2011}. This shows that the extracted
intrinsic carrier density is realistic and that THz radiation is
an excellent probe for the free carrier conductivity. We note that
the substrate thickness of 200 $\mu$m results in a small
additional peak in the pump-probe delay traces at a delay time of
$\sim$3 ps due to reflection of the THz wave. We have confirmed
that this peak is absent when using a thicker substrate.
\\

In our measurements we measure the pump-induced change of the
conductivity of the free carriers, i.e.\ the THz
photoconductivity. In order to make quantitative comparisons
between the THz photoconductivity induced by photo-excitation with
different photon energies, we need to accurately know the number
of absorbed photons at each pump wavelength. We obtain this number
through the measurement of the fluence (the energy of the pump
beam per unit of area) and through the graphene absorption at each
pump wavelength. We determined the graphene absorption using a
standard double beam spectrometer (Perkin Elmer Lambda 950), where
we placed an identical quartz substrate without graphene in the
reference path. The absorption was determined from 200 nm up to
2.2 $\mu$m (see Fig.\ \ref{F_SOM_abs_raman}C), above which the
substrate starts absorbing strongly. The change in reflection due
to the presence of graphene is negligibly small ($<$0.1$\%$)
\cite{Nair2008}. We find relatively flat absorption that is
slightly lower than the universal value of 2.3$\%$
\cite{Nair2008}, down to a wavelength of $\sim$500 nm and
increased absorption due to excitonic effects that peak around 250
nm \cite{Mak2011}.
\\

\section{Differential Transmission, THz Conductivity, and THz
photoconductivity}

The transmission, $T(\omega)$, of light at a frequency $\omega$
through a single graphene layer and substrate stack (in the thin
film approximation) is related to its optical conductivity,
$\sigma (\omega)$ via \cite{orenstein,Strait2011} \be
\frac{T}{T_{\rm sub}} = \Big[1+ \frac{\sigma(\omega) Z_0}{n_s + 1}
\Big]^{-1}, \ee where $T_{\rm sub}$ is the transmission through
the substrate only (no graphene), $Z_0 = \sqrt{\mu_0/\epsilon_0}$
is the impedance of free space, and $n_s$ is the index of
refraction of the substrate. Throughout this text, we only
consider the {\it real} part of the optical conductivity which we
will denote by $\sigma$. In pump-probe experiments, It is natural
to consider the change in transmission that is due to the pumping
of the sample. The difference between pump on and pump off
transmission, $\Delta T = T - T_0$ where $T_0$ corresponds to the
transmission with pump off, is also related to the change in
optical conductivity of the graphene sample due to optical
pumping. This gives \be \frac{\Delta T}{T_0} = \frac{ 1+
[\sigma_0(\omega) Z_0]/[n_s + 1]}{ 1+ [\sigma_1(\omega) Z_0]/[n_s
+ 1]} -1 \approx - \Delta \sigma(\omega) \frac{Z_0}{n_s+1}, \ee
where $\sigma_0(\omega)$ is the optical conductivity with pump
off, $\Delta \sigma = \sigma_1(\omega) - \sigma_0(\omega)$ and in
the last line we have used $[\sigma(\omega) Z_0]/[n_s + 1] \ll 1$
and expanded the denominator. $\sigma_1$ is the conductivity when
the pump is on. For $\omega$ in the ranges of $1 \, {\rm THz}$
this is referred to as the Terahertz conductivity and probes
intra-band scattering events as $\hbar \omega \ll E_F$.

The optical conductivity depends directly on the energy-dependent
carrier distribution and gives a different value according to how
far away from equilibrium the carrier distribution is. To see
this, we use the kinetic equation to derive the
conductivity\cite{grosso}
\be
\partial_t n_{\vec k}+e\vec E\nabla_{\vec k}n_{\vec k}=\sum_{\vec k'}W_{\vec k',\vec k}\lb
(1-n_{\vec k})n_{\vec k'}- (1-n_{\vec k'})n_{\vec k}\rb
=\sum_{\vec k'}W_{\vec k',\vec k}\lp n_{\vec k'}-n_{\vec k}\rp \ee
The perturbation of the non-equilibrium distribution due to an
external electric field, $n_{\vec k}=\bar n_{\vec k}+\delta
n_{\vec k}$,
can be found by the relaxation time approximation where we 
linearize the right hand side in $\delta n_{\vec k}$ giving $\sum_{\vec k'}W_{\vec k',\vec k}\lp n_{\vec k'}-n_{\vec k}\rp \approx -\delta n_{\vec k}/ \tau_{\rm tr}(\vec k)$ where $\frac1{\tau_{\rm tr}(\vec k)}=\sum_{\vec k'}W_{\vec k',\vec k}(1-\cos\theta)$. 
%
%
At finite frequency the carrier distribution response to an AC
field $\vec E\propto e^{-i\omega t}$ gives a (real part of)
conductivity
\be \label{eq:conductivity} \sigma (\omega) =-\sum_{\vec
k}\frac{\tau_{\rm tr}(\epsilon_{\vec k})}{1+ [\omega\tau_{\rm
tr}(\epsilon_{\vec k})]^2}e^2v_{\vec k}\nabla_{\vec k} n_{\vec k}
\ee
where we have used the formula for current $\vec j=\sum_{\vec k}ev_{\vec k}\delta n_{\vec k}$ and picked out the $\omega$ response. 

As described in the main text upon pumping by optical pulse, a
narrow band of photo-excited carriers is created. These
photo-excited carriers can relax energetically in two primary ways
depicted in Fig. 1 of the main text : (i) carrier-carrier
scattering with carriers in the Fermi sea or by emission of
optical phonons. Both processes can lead to a change in the THz
conductivity : termed THz photoconductivity. The former excites
the carrier distribution about the Fermi surface leading to a
hotter carrier distribution and a change in the THz conductivity.
Because the transport time increases with carrier energy
\cite{nomura,ando,dassarmareview}, the change in the real part of
conductivity (measured in transmission) is negative. The latter
process produces optical phonons which can act as scattering
centers for carriers in the Fermi sea, increasing the resistivity
of the sample.

In the following we consider both processes. Accounting for fast
thermalization of the secondary carriers, the net change in the
THz photoconductivity can be expressed as $-2 N_{\rm photon}\Big(
a \mathcal{J}_{\rm el-el} + b \mathcal{J}_{\rm el-ph}\Big)$, with
the energy relaxation of the photoexcited carriers described in a
general form via $d{\epsilon}/dt=-\mathcal{J}_{\rm
el-el}(\epsilon)  - \mathcal{J}_{\rm el-ph}(\epsilon)$, where
$\epsilon$ is the photo-excited carrier energy and
$\mathcal{J}_{\rm el-el} $ and $\mathcal{J}_{\rm el-ph}$ represent
the energy relaxation rates for processes (i) and (ii). Here  $a$
is the coefficient governing the change in THz conductivity from
the capturing of energy by the electronic system (see below), and
$b$ is the coefficient governing the change in THz conductivity
from the emission of optical phonons that act as scattering
centers (see below). As a result, the THz photoconductivity from
these two processes is

\be \label{eq:photocontot} \Delta \sigma = -2 N_{\rm photon}
\int_{t_0}^{t_0+\tau_{\rm peak}} dt ( a\mathcal{J}_{\rm el-el} + b
\mathcal{J}_{\rm el-ph}) = -2 N_{\rm photon} \int^{hf/2}_{0} dE
\frac{ a\mathcal{J}_{\rm el-el} + b\mathcal{J}_{\rm el-ph}
}{\mathcal{J}_{\rm el-el}  + \mathcal{J}_{\rm el-ph} } , \ee

To get a gauge of which processes contribute most to the
photoconductivity, we note that for a photon energy of $hf = 3.1
\, {\rm eV}$, and $N_{\rm photon} = 10^{11} \, {\rm cm}^{-2}$ our
experiments found a transmission change of $0.75 \%$ (Fig. 1C of
main text) which corresponds to a photoconductivity of $ \Delta
\sigma \approx  -1.5 e^2/h$.  Using values of $b$, $E_F=0.17 \,
{\rm eV}$ in our samples, $\mathcal{J}_{\rm el-el}$, and
$\mathcal{J}_{\rm el-ph}$ calculated below, we find that the
optical phonon scattering contribution to photoconductivity is
$\Delta \sigma_{\rm ph} \approx - 0.02\, e^2/h$ about $75$ times
smaller than what is observed. Hence, we conclude that the
scattering off emitted optical phonons produce negligibly small
contribution to the photoconductivity. As a result, we set $b=0$
in our analysis of our data (see Eq. 1, Fig. 2 of main text and
discussion below).

\section{Electronic temperature model for $a$} \label{S:a}

We consider the photoconductivity due to carrier-photoexcited
carrier scattering. This process excites the carrier distribution
about the Fermi surface. Because carrier-carrier scattering is
fast on the order of $10$s of $\rm {fs}$ \cite{george,song12} -
shorter than the initial rise time $\sim 200\, {\rm fs}$ measured
in our pump-probe experiment  (see Fig.\ 1C of the main text) -
the excited carrier distribution thermalizes quickly and can be
described by an increased effective electronic temperature, $T$.
Thermal equilibration with the lattice in graphene is slow
\cite{bistritzer09,wong09,song12a}, and in our samples takes $\sim
1.4\, {\rm ps}$ (see Fig.\ 1B  of main text). This separation of
time scales allows us to treat the electronic system as out of
equilibrium with the lattice.

In the degenerate limit, $E_F \gg k_BT$, we can use the Sommerfeld
expansion in Eq. \ref{eq:conductivity} giving a (real part of) THz
conductivity \be \label{eq:sommerfeld} \sigma(\omega) \approx
\sigma(\omega)_{T=0} + \frac{\pi^2}{6}\nu(E_F) k_B^2 T^2
\frac{\partial^2F(\epsilon)}{\partial
\epsilon^2}\Bigg|_{\epsilon=E_F} , \quad F(\epsilon)= e^2 v^2
\frac{\tau_{\rm tr}(\epsilon)}{1+\omega^2[\tau_{\rm
tr}(\epsilon)]^2} \ee where $E_F$ is the Fermi energy ,
$\nu(\epsilon)$ is the total density of states in graphene, and $v$ is the Fermi velocity. 
We have kept the total carrier density, $n$, constant by
accounting for changes in chemical potential as a function of
temperature $\mu \approx E_F - \frac{\pi^2}{6} k_B^2 T^2 / E_F$.
Here we have neglected the temperature dependence of the transport
time, $\tau_{\rm tr}(\epsilon)$, since we estimate that $k_B\Delta
T < E_F$. We note that these, in principle, can provide additional
terms to Eq. \ref{eq:sommerfeld} and can be included in a more
sophisticated analysis.

Therefore, the hot carrier contribution to the photoconductivity
arising from an increase in temperature of the
carrier-distribution, $\Delta \sigma_{\rm el} = \sigma_1 -
\sigma_0$ is \be \Delta \sigma_{\rm el} =  \Big( k_B^2 T_1^2 -
k_B^2 T_0^2\Big) \frac{\pi^2}{6}
\nu(E_F)\frac{\partial^2F(\epsilon)}{\partial
\epsilon^2}\Bigg|_{\epsilon=E_F} \ee where $1,0$ subscripts
indicate pump on and off respectively.

We note parenthetically that the differential transmission in our
set-up is related to the the photoconductivity above but with a
sign that is opposite to $\frac{
\partial^2 F(\epsilon)}{\partial \epsilon^2}$. For graphene,
$\frac{ \partial^2 F(\epsilon)}{\partial \epsilon^2} < 0$ for
small probing frequencies $\omega \sim {\rm THz}$ because
$\tau_{\rm tr} \propto \epsilon$ \cite{nomura, ando,
dassarmareview}. As a result, our observed positive differential
transmission (or equivalently negative photoconductivity) is
consistent with a hotter carrier distribution. On the other hand,
if carrier density had increased (for example, by exciting
electrons from the valence band), this would have resulted in a
positive photoconductivity (which we did not observe).

We can relate $\Delta \sigma_{\rm el}$ to the change in heat in
the system. In the degenerate limit, the electronic heat capacity
is $C = \alpha T$ where $\alpha = \frac{\pi^2}{3} \nu(E_F) k_B^2$.
Hence, the photoconductivity is \be \Delta \sigma_{\rm el} =
\frac{k_B^2\Delta Q}{\alpha}\frac{\pi^2}{3}
\nu(E_F)\frac{\partial^2F(\epsilon)}{\partial \epsilon^2} ,\quad
\Delta Q = \frac{\alpha}{2} (T_1^2- T_0^2). \ee where $\Delta Q =
Q_1 - Q_0$. This means that $\Delta \sigma_{\rm el}$ is a direct
measure of the amount of heat captured by the electronic system.
Since the amount of heat absorbed by the carriers in the Fermi sea
comes directly from the energy-loss of the photo-excited carrier,
the rate of heat entering the Fermi sea is \be \frac{d\Delta
Q}{dt} = \mathcal{J}_{\rm el-el} N_{\rm photon} \ee where $-
\mathcal{J}_{\rm el-el}$ is the energy-loss rate of the
photo-excited carrier due to carrier-carrier scattering between
the photo-excited carrier and the carriers in the Fermi sea and
$N_{\rm photon}$ is the absorbed photon density (ie. density of
initial photo-excited carriers). We have neglected thermal
equilibration with the lattice as the time scales for cooling to
the lattice $\sim 2\, {\rm ps}$ are far longer than the time
scales of energy-loss of the photo-excited carrier $\sim 40\, {\rm
fs}$ \cite{song12, george}.

As a result, the photoconductivity from the relaxation of the
photo-excited carriers produces the first term in Eq.
\ref{eq:photocontot} with \be a = - \frac{
\partial^2F(\epsilon)}{\partial \epsilon^2} \ee Here $a$ is a
positive quantity. As seen above, the photo-conductivity (and
hence the differential transmission) is directly sensitive to the
energy relaxation dynamics of the photo-excited carrier and
$\Delta \sigma_{\rm el}$ a unique probe of the amount of energy
that gets captured by the electronic system.

In the following, we estimate $a$ from this model. Since
conductivity in graphene for high doping depends linearly on
density $n \propto \mu ^2$ \cite{dassarmareview}, the einstein
relation, $\sigma = e^2 v^2 \nu(\mu) \tau(\mu) / 2$ tells us that
$\tau (\epsilon) = \alpha \epsilon$, where $\alpha$ is a
proportionality constant. The factor of $2$ in the einstein
relation comes from dimensionality.This is consistent with
carriers scattering off Coulomb disorder in graphene
\cite{nomura,ando,dassarmareview}. We measured a transport time at
$\epsilon = \mu$ of $20 \, {\rm fs}$ (see above section) and a
doping corresponding to $E_F = 170 \, {\rm meV}$ (see Raman
Spectroscopy above). Therefore, we infer that in our sample
$\alpha \approx 117 \, {\rm fs} \, {\rm eV}^{-1}$.

Differentiating $F$, and using  $E_F = 0.17 \, {\rm eV}$ and a
typical THz frequency that was used $\omega = 2\pi \, {\rm THz}$
(corresponding to $f \sim 1\, {\rm THz}$)  we obtain


\begin{align}
a \approx  2.68 \times 10^{-12}  \times \frac{e^2}{h} {\rm
cm}^{2}{\rm eV}^{-1}
\end{align}

This value for $a$ gives a photoconductivity of $\Delta \sigma =
-0.83 e^2/h$ for a photon energy of 3.1 eV and absorbed photon
density of $N_{\rm photon} = 10^{11}$ cm$^{-2}$, using the
theoretical values of $\mathcal{J}_{\rm el-el}$ and
$\mathcal{J}_{\rm el-ph}$ obtained below. Experimentally, we
measure a differential transmission signal of $\Delta T/T_0 \sim$
0.75\% for these conditions, which corresponds to $\Delta \sigma =
-1.5 e^2/h$, in very reasonable agreement with the theoretical
result. A comparison of $a$ with $b$ (estimated below) yields $a/b
\approx 5$ for photoexcited carrier energy of $\epsilon =1 \, {\rm
eV}$ and $\mu=0.17 \, {\rm eV}$.

\section{Optical phonon scattering contribution to
photoconductivity, $b$}

We now consider the contribution to photoconductivity that comes
from the emission of optical phonons. Optical phonons can act as
scattering centers for carriers in the Fermi sea and alter the
mobility of the graphene sample. Hence, a larger population of
optical phonons increases the resistivity of the sample, $\Delta
\rho_{\rm ph}$, and can lead to a negative photoconductivity,
$\Delta \sigma_{\rm ph} = -\sigma_0^2 \Delta \rho_{\rm ph}$. Here,
$\sigma_0$ is the conductivity without pump excitation. In
particular, the optical phonon limited resistivity is
\cite{avouris} \be \Delta \rho_{\rm ph} = \frac{2\pi D_{\rm op}^2
\hbar N(\omega_0)}{e^2 \rho_m v^2 \omega_0} \label{eq:opticalrho}
\ee

where $D_{\rm op}$ is an effective electron-optical phonon
coupling constant \cite{avouris}, $\rho_m $ is the mass density of
graphene, $\omega_0$ is the optical phonon energy of graphene  and
$N(\omega_0)$ is the occupancy of optical phonons in graphene.

To estimate the occupancy $N(\omega_0$ we note that the optical
phonons that are emitted from the energy-relaxation of the
photo-excited carrier similarly only have momenta $< \epsilon /
(\hbar v)$ where $\epsilon$ is the energy of the photo-excited
carrier.
%
%
Hence, the (conditional) probability density, $p$, of emission of
an optical phonon with wavevector $\vec q$ given that an optical
phonon was emitted by the energy relaxation of a high energy
photo-excited carrier with wavevector $\vec{k}$ is \be dp =
\frac{d\theta}{\pi} (1- {\rm cos} \theta) =  \frac{1}{\pi}
\frac{1}{2|\vec{k}|} \sqrt{\frac{2|\vec{k}| - |\vec q|}{2|\vec{k}|
+ |\vec q|}} dq, \quad |\vec{q}| = 2 |\vec{k}| {\rm cos} \theta
\ee where $\theta$ is the angle between $\vec q$ and $-\vec{k}$,
and the photo-excited carrier's energy is $\epsilon = v\hbar |\vec
k|$. Here we have used the factor $1- {\rm cos} \theta$ to account
for the coherence factor in the scattering matrix element. Since
optical phonon scattering with carriers in the Fermi sea is mainly
limited $k_F$ far smaller than the Brillouin zone we are
interested in the fraction of emitted optical phonons that lie
within the circle of radius $2k_F$. The fraction of optical
phonons emitted that lie within the circle of radius $2k_F$ is \be
p(|\vec q|< 2k_F) = \int_0^{2k_F} \frac{1}{\pi}
\frac{1}{2|\vec{k}|} \sqrt{\frac{2|\vec{k}| - |\vec q|}{2|\vec{k}|
+ |\vec q|}} dq \approx \frac{1}{\pi} \frac{k_F}{|\vec{k}|} \ee As
a result over the course of the relaxation of photo-excited
carriers, the number density of optical phonons emitted within
this circle of radius $2k_F$, is \be \label{eq:occupancy}
 \sum_{|\vec q | < 2 k_F}
N(|\vec q|)= N_{\rm photon} \int \frac{\mathcal{J}_{\rm el-ph}
/\omega_0}{\mathcal{J}_{\rm el-ph}  + \mathcal{J}_{\rm el-el} }
\times \frac{1}{\pi} \frac{k_F}{|\vec{k}|} d\epsilon \ee where
$N(|\vec q|)$ is the occupancy of optical phonons emitted. On
average, each optical phonon in this circle of radius $2k_F$
contributes the same amount to the transport scattering rate
between optical phonons and carriers in the Fermi surface. Hence,
the average occupancy within this circle, $N_{\rm av}^{|\vec q|<
2k_F} = \big[\sum_{|\vec q | < 2 k_F} N(|\vec q|) \big] /
({\sum_{|\vec q|< 2k_F}}  )$, allows us to estimate the change in
resistivity coming from these optical phonons via Eq.
\ref{eq:opticalrho}. Therefore, noting that $\sum_{|\vec q|< 2k_F}
= \frac{\pi (2k_F)^2}{(2\pi)^2} = k_F^2/\pi$ and using Eq.
\ref{eq:occupancy} and Eq. \ref{eq:opticalrho}, we find that
emission of optical phonons contributes a photoconductivity
described by the second term of Eq. \ref{eq:photocontot} with
\be b = \frac{2\pi \sigma_0^2 \hbar^2 D_{\rm op}^2 }{e^2 \rho_m v
\omega_0^2 k_F \epsilon} \ee

We can estimate the value of $b$. Taking $D_{\rm op} = 22.4\, {\rm
eV} \, {\AA}^{-1}$ \cite{avouris}, $\rho_m  = 7.6 \times 10^{-11}
\, {\rm kg} \,{\rm cm}^{-2}$, $\omega_0 = 0.2 \, {\rm eV}$, and
$\mu = 0.17 {\rm eV}$ we obtain \be b \approx \frac{5.74\times
10^{-13}}{(\mu \, [ {\rm eV}]/0.2)\times(\epsilon [{\rm eV}]) }
\times \frac{e^2}{h} {\rm cm}^2 {\rm eV}^{-1}, \ee where we have
used the measured conductivity $\sigma_0 \approx 10 (e^2/h)$ (see
section \ref{S:sample_characterization}).

This value for $b$ gives a photoconductivity of $\Delta \sigma =
-0.02 e^2/h$ for a photon energy of 3.1 eV and absorbed photon
density of $N_{\rm photon} = 10^{11}$ cm$^{-2}$. Experimentally,
we found $\Delta \sigma = -1.5 e^2/h$, which is significantly
higher than the signal due to phonon emission only. We showed in
Section \ref{S:a} that the signal due to hot carriers is much
larger and closer to the experimentally observed photoconductivity
signal.

\section{Energy-relaxation of photo-excited carriers,
$-\mathcal{J}$}

The energy relaxation of a photo-excited carrier occurs via two
processes: (i) scattering between the photo-excited carrier and
the carriers in the Fermi sea (see Fig. 1B main text) allows the
energy to be transferred to the Fermi sea to create a hot carrier
distribution, and (ii) the emission of optical phonons. Because of
the large number of carriers in doped graphene, the former process
can be very efficient \cite{song12}.

\begin{figure} [h!!!!!]
   \centering
   \includegraphics [scale=0.8]{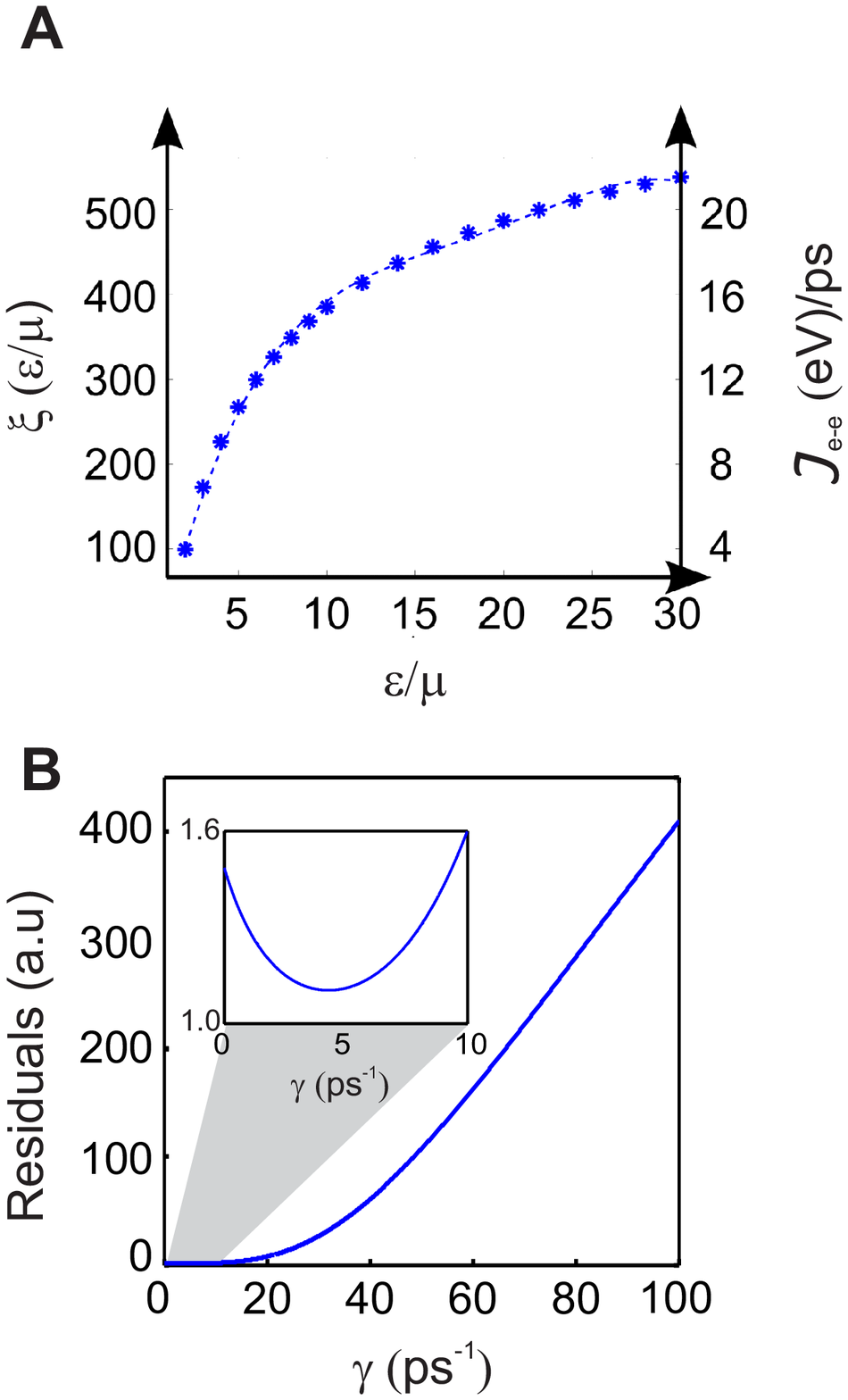}
   \caption{\textbf{A)}
    (Left ordinate) Starred points denote numerically evaluation of scaling function
   $\xi(\epsilon/\mu)$ using Ref. \cite{song12}. Dashed lines are a 4th order polynomial fit
   of the numerical evaluation within this range: $\xi(x) = -3.1081 \times 10^{-3} x^4 +
   0.24031x^3 - 6.8417x^2 + 92.622x-59.383$ where $x= \epsilon/\mu$. (Right ordinate)
   Magnitude of $\mathcal{J}_{\rm el-el}$ for $\mu=0.2 \, {\rm eV}$ using Eq. \ref{eq:jelel}.
    \textbf{B)} The squared residuals as a function of $\gamma$, obtained by fitting the theoretical results for
    $\frac {-\Delta \sigma} {N_{\rm ph}} (\epsilon)$ to the
    experimentally obtained $\frac{\Delta T}{T_0 N_{\rm ph}} (\epsilon) \propto - \frac{\Delta \sigma} {N_{\rm ph}} (\epsilon)$.
    The fit is done for a range of values for $\gamma$, which
    reflects the branching ratio, and with one free fitting
    parameter that is just a scaling factor. The best fit is
    obtained for $\gamma = $4 ps$^{-1}$, and the residuals increase strongly for
     large values of $\gamma$. }
   \label{SOM-energyrelaxation}
\end{figure}

Additionally, the energy relaxation via this channel occurs in
relatively small steps of order $\mu$ the energy relaxation rate
from impact excitation events, $\mathcal{J}_{\rm el-el}$
\cite{song12}, is

\be \mathcal{J}_{\rm el-el} (\epsilon) = (\mu [eV])^2
\xi(\epsilon/\mu)\,{\rm eV}/{\rm ps} \label{eq:jelel} \ee

where $\xi(\epsilon,\mu)$ is plotted in Fig.\
\ref{SOM-energyrelaxation} using a numerical integration detailed
in Ref. \cite{song12} (starred points in Fig.
\ref{SOM-energyrelaxation}). We subsequently used a 4th order
polynomial to interpolate between the points (dashed line in Fig.
\ref{SOM-energyrelaxation}). For our doping, $\mu=0.17 \, {\rm
eV}$ we find an efficient relaxation rate that for typical carrier
energies are a couple of ${\rm eV}/{\rm ps}$ indicating that the
relaxation of photoexcited carriers with typical energy of
$\epsilon = 1\, {\rm eV}$ via $\mathcal{J}_{\rm el-el}$ occurs
with a few hundred femtoseconds. This is consistent with our
experimental observation of a rise time of around $\tau_{\rm peak}
\sim 200 \, {\rm fs}$ (see main text Fig. 1).


An alternative channel for energy relaxation of the photo-excited
carrier occurs through the emission of optical phonons and gives
an energy relaxation rate of  $-\mathcal{J}_{\rm el-ph}$. The
transition rate of this process \cite{tse08} can be described by
Fermi's golden rule

\be W_{\vec{k}',\vec{k}}^{\rm el-ph}  = \frac{2\pi
N}{\hbar}\sum_{\vec q} |M(\vec{k}',\vec{k})|^2 \delta
\big(\Delta\epsilon_{\vec{k}', \vec{k}}  + \omega_\vec{q}  \big)
\delta_{\vec{k}', \vec{k} + \vec q} (N({\omega_\vec{q}}) + 1) \ee

where $\Delta\epsilon_{\vec{k}', \vec{k}} = \epsilon_{\vec k'} -
\epsilon_{\vec k}$, $\omega_\vec{q}=\omega_0 = 200\, {\rm meV}$ is
the optical phonon dispersion relation, and $N({\omega_{\vec q}})$
is a Bose function. Here $\vec k$ is the initial momentum of the
photo-excited electron, $\vec k'$ is the momentum it gets
scattered into, and $\vec q$ is the momentum of the optical
phonon. The electron-phonon matrix element $M(\vec{k}',\vec{k})$
\cite{bistritzer09} is

\be \label{eq:matrixelement} |M(\vec{k}',\vec{k})|^2 = g_0^2
F_{\vec{k}, \vec{k}' }, \quad g_0 = \frac{2\hbar^2v}{\sqrt{2\rho_m
\omega_0 a^4}} \ee

where $ F_{\vec{k}, \vec{k}' }$ is the coherence factor for
graphene, $g_0$ is the electron-optical phonon coupling constant,
$a$ is the distance between nearest neighbor carbon atoms, and
$\rho_m$ is the mass density of graphene. The energy-loss rate of
the photo-excited carrier at energy $\epsilon$ due to the emission
of an optical phonon is
\be \mathcal{J}_{\rm el-ph}(\epsilon) = \sum_{\vec{k'}}
W_{\vec{k}', \vec{k}}^{\rm el-ph} (\epsilon_k' - \epsilon)\big[ 1-
f(\epsilon_\vec{k'})\big] \ee Integrating over $\vec q$ and $\vec
k'$ we obtain
\be\label{eq:jelph} \mathcal{J}_{\rm el-ph}(\epsilon) = \frac{\pi
N}{\hbar} \omega_0 g_0^2 \big[ 1- f(\epsilon - \omega_0)\big]
(N(\omega_0)+1)\nu(\epsilon - \omega_0) =  \gamma (\epsilon - \omega_0) \Theta(\epsilon - E_F - \omega_0),
\ee
where $\nu(\epsilon)=\epsilon/(2\pi v^2\hbar^2)$ is the electron
density of states in graphene and we have approximated
$(N(\omega_0)+1) \approx 1$ and $1- f(\epsilon - \omega_0) \approx
\Theta(\epsilon - \omega_0- E_F)$. $\gamma$ is a constant
determined by the electron-optical phonon coupling. Hence,
$\mathcal{J}_{\rm el-ph}(\epsilon)$ varies linearly with the
photo-excited carrier energy $\epsilon > \omega_0$ and vanishes
for $\epsilon<\omega_0$. Because the electron-phonon coupling with
optical phonon is a constant, this result is to be expected from
the increased phase space to scatter into at higher photo-excited
carrier energy.

Using $\rho_m=7.6 \times 10^{-11} \, {\rm kg} \, {\rm cm}^{-2}$
and $a = 1.42 \AA$ in Eq. \ref{eq:matrixelement} above yields $
\gamma \approx \, 1.36 {\rm ps}^{-1}. \label{eq:gammatheory} $
\\

\section{Comparison of data with model}

These theoretical results for $\mathcal{J}_{\rm el-el}$ and
$\mathcal{J}_{\rm el-ph}$ lead to a branching ratio between
electron-electron scattering and electron-optical phonon
scattering that we wish to compare to the experimental data.
Therefore we take the experimental results for the signal
normalized by absorbed photon density (Fig.\ 2A of the main text),
which is directly proportional to $- \frac {\Delta\sigma} {N_{\rm
ph}}$, and compare these with the model through Eq.\ 1 of the main
text. We fix $\mathcal{J}_{\rm el-el}$ to the theoretical result
and use $\mathcal{J}_{\rm el-ph} = \gamma \Theta(\epsilon - E_F -
\omega_0)$, where we use the electron-phonon coupling constant
$\gamma$ as the parameter that determines the experimental
branching ratio. The parameter $\gamma$ reflects how $- \frac
{\Delta\sigma} {N_{\rm ph}}$ depends on carrier energy (or $hf$):
A low value of $\gamma$ corresponds to an energy cascade that is
dominated by el-el scattering and results in linear scaling with
carrier energy, whereas a high value corresponds to el-ph
dominating, which leads to the signal bending down; large $\gamma$
leads to a non-linearity in $\Delta \sigma$ setting in at a
low carrier energy. \\

We repeat the fit routine for a range of values for $\gamma$ and
include one fit parameter that determines the amplitude of the
signal, thus serving as a mere scaling factor. In Fig.\ 2A we use
the amplitude obtained from the best fit for all three curves,
such that all curves for different $\gamma$ pass through the same
data point at low excitation energy (0.56 eV). We do not include
the lowest (0.16 eV) and highest (4.65 eV) photon energies. In the
former, the model does not cover this region and the absorbed
fluence is not known very precisely. In the latter,
 the diffuser could not be used, which leads to possible
over-estimation of the absorbed photon density (in case of hot
spots in the beam profile). The fit results for $\gamma$ = 1, 4
and 10 ps$^{-1}$ are shown in Fig.\ 2A of the main text. Here we
show the residuals (squared difference between experimental data
points and model) for a range of $\gamma$'s, clearly showing a
minimum around 4 ps$^{-1}$ and strong deviation especially at
large values of $\gamma$ (see Fig. S2 B). This clearly shows that
our data is incompatible with a phonon-dominated energy cascade.
The best fitting value is somewhat larger than the theoretically
obtained one (1.36 ps$^{-1}$), but the theoretical value still
lies well within reach of the experimental data. It is also
conceivable that the data leads to an overestimation of $\gamma$
(and thus an underestimation of the contribution of el-el
scattering) due to a nonlinear (saturation) response of the signal
at high photon fluences and high photon energy. This means that we
extract a conservative branching ratio.
\\

\end{document}